 \documentclass[preprint2]{aastex}

\shorttitle{Activity and rotation of  Kepler-17}
\shortauthors{Valio et al.}

\begin{document}

\title{Activity and rotation of Kepler-17}

\author{Adriana Valio, Raissa Estrela, Yuri Netto}
\affil{CRAAM, Mackenzie Presbyterian University, Rua da Consolacao, 896, Sao Paulo , Brazil}
\email{avalio@craam.mackenzie.br}


\author{J. P. Bravo, and J. R. de Medeiros}
\affil{Universidade Federal do Rio Grande do Norte, Natal, Brazil}

\begin{abstract}
Magnetic activity on stars manifests itself in the form of dark spots on the stellar surface, that cause modulation of a few percent in the light curve of the star as it rotates. When a planet eclipses its host star, it might cross in front of one of these spots creating a ``bump" in the transit light curve. By modelling these spot signatures, it is possible to determine the physical properties of the spots such as size, temperature, and location. In turn, the monitoring of the spots longitude provides estimates of the stellar rotation and differential rotation. This technique was applied to the star Kepler-17, a solar--type star orbited by a hot Jupiter. The model yields the following spot characteristics: average radius of $49 \pm 10$ Mm, temperatures of $5100 \pm 300$ K, and surface area coverage of $6 \pm 4$ \%. The rotation period at the transit latitude, $-5^\circ$, occulted by the planet was found to be $11.92 \pm 0.05$ d, slightly smaller than the out--of--transit average period of $12.4 \pm 0.1$ d. Adopting a solar like differential rotation, we estimated the differential rotation of Kepler-17 to be $\Delta\Omega = 0.041 \pm 0.005$ rd/d, which is close to the solar value of 0.050 rd/d, and a relative differential rotation of  $\Delta\Omega/\Omega=8.0 \pm 0.9$ \%. Since Kepler-17 is much more active than our Sun, it appears that for this star larger rotation rate is more effective in the generation of magnetic fields than shear.

\end{abstract}

\keywords{Stellar activity, starspots, stellar rotation, stellar differential rotation}

\section{Introduction}\label{Intro}

In the Sun, spots can be seen throughout the disk, at different longitudes and latitudes that change over the course of the magnetic cycle. By analogy with sunspots, spot activity is also observed on all main-sequence late-type stars and is believed to be generated by the interaction between the magnetic field, convection, and differential rotation  \citep{Lammer15}. As the star rotates, starspots cause modulation in the light curve of the star, varying its brightness. This enables the measurement of the rotational period of the star, as well as characterisation of the brightness and relative size of starspots.

Spot modelling using disk-integrated light modulation, however, identifies only a few large spots in the stellar surface. 
The analysis of planetary transits appears as a better approach because they can show detectable variations in the light curve caused by the passage of the planet in front of a solar-like spot on the stellar surface \citep{Silva03}. These spots can produce different effects in the transit shape. For example, spots that are not occulted by the planet will produce a deeper transit, while spots occulted by the transit provoke an increase in the luminosity flux \citep{Pont13}. These same effects may also affect the transmission spectroscopy measurements of the transit \citep{Oshagh14}. To avoid any influence of the spots in the radius ratio of the transmission spectroscopy, \cite{Sing11} corrected the HD 189733 data by calculating the variation in the flux caused by the spot and also estimating its temperature. 

Making use of the activity signatures in planetary transits, \cite{Silva03} developed a method that allows the detection of spots as small as $\sim$0.1 planetary radius. Moreover, this approach also infers the properties of individual starspots along the occulted band, such as size, intensity and position. Such technique was previously applied to HD 209458 \citep{Silva03}, to the active star CoRoT-2 \citep{SilvaValio10,SilvaValio11}, to the characterisation of the magnetic activity cycle of stars Kepler-17 and Kepler-63 \citep{Estrela16}, and recently to estimate the differential rotation of Kepler-63 \citep{Netto16}. Although not explored in this work, the analysis of the same spot in different transits can also provide information about the spin-orbit alignment of the star-planet system \citep{Sanchis11,Sanchis12}.

\begin{figure}[h]
\plotone{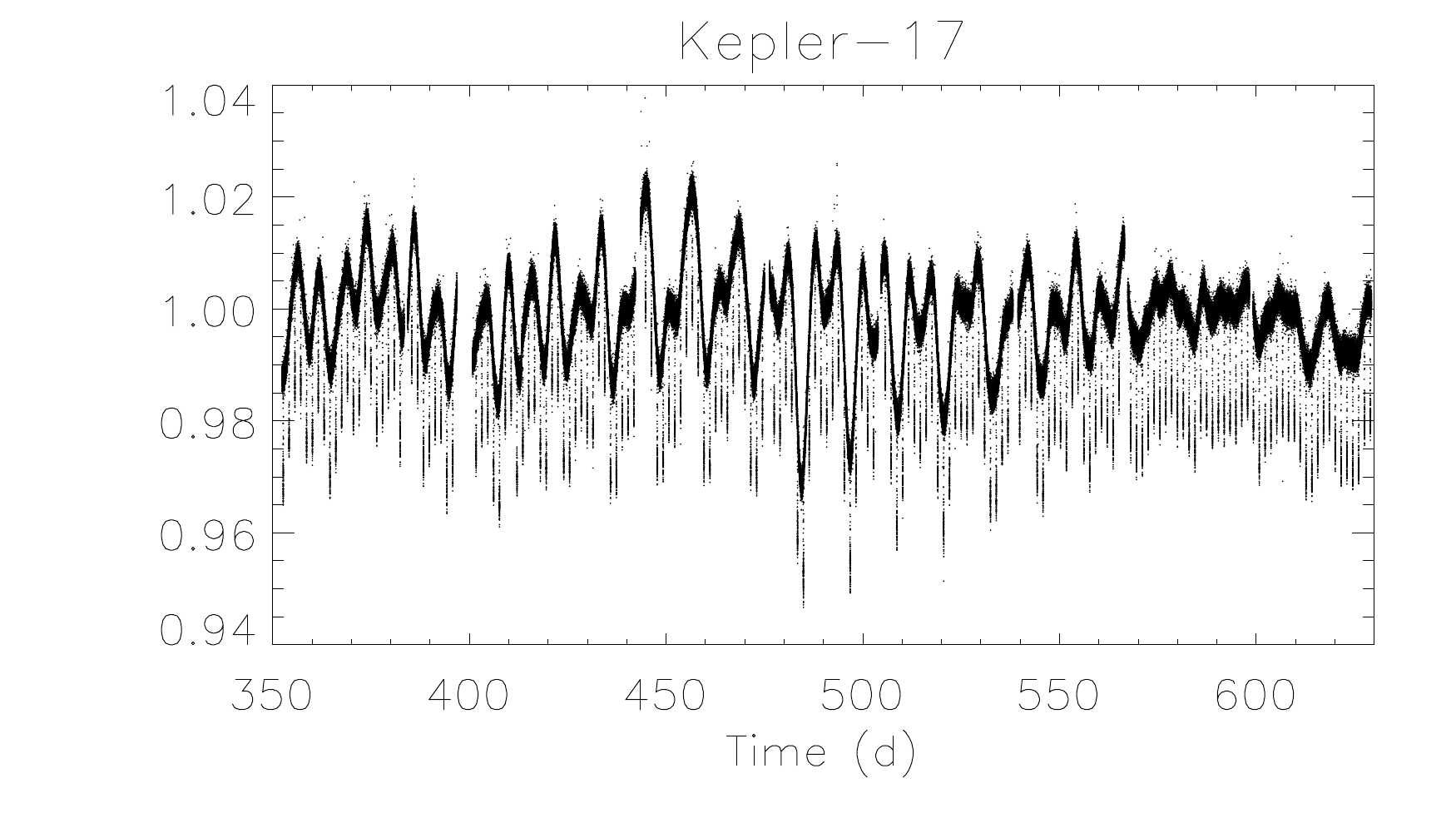}
\caption{Extract of Kepler-17 light curve showing the $\sim 4\%$ photometric modulation due to the presence of spots on the stellar surface that rotate in and out of view.}
\label{lcurve}
\end{figure}

The solar-type star analysed here, Kepler-17, is a prominent candidate to study stellar magnetic activity, as its light curve shows rotational modulations caused by the presence of starspots, with an amplitude variation of 4$\%$ (Figure~\ref{lcurve}). The Sun, for comparison, presents a modulation of only about 0.1\%. Kepler-17 has a Hot Jupiter in its orbit \citep{Desert11} that occults the spots during the transit producing a detectable signal in the transits profiles. The active regions of this star were previously studied by \cite{Bonomo12a} who estimated a spotted area about 10 to 15 times larger than those of the Sun, and found evidence of a solar-like latitudinal differential rotation, later confirmed by \cite{Davenport15}, but its value was not estimated.  

A thorough investigation of differential rotation is crucial to understand the magnetic dynamo of the star, which is responsible for the generation of the stellar magnetic field. Therefore, we propose to use the transit technique to investigate the characteristics of the spots in Kepler-17 along with the rotation behaviour of the star, by detecting the same spots in later transits. Consequently, from the study of  starspots we are able to measure the rate of differential rotation of the star, by considering a rotation profile similar to the Sun. 

This paper is organised as follows. The following section gives an overview about the observations of the Kepler-17 system. Section~\ref{Model} provides a detailed explanation of the method used in this work, whereas Section~\ref{Spots} describes the physical parameters of the modelled spots. Analysis of the longitude maps of the spots used to determine the rotation period at the transit latitude and thus estimate the differential rotation period is explained in Section~\ref{Res}. Finally, the last section presents our conclusions.

\section{Observations of Kepler-17}\label{Obs}

Kepler-17 is a G2V star ($T_{eff}=5781$ K), the same spectral type as our Sun,  with a mass of $1.16 ~M_\odot$ and radius of $1.05 ~R_\odot$, that was observed by the Kepler satellite for almost 4 years. It is a young star with less than 1.78 Gyr \citep{Bonomo12b}, that rotates with an average period of $12.4 \pm 0.1$ d.

\begin{table}[h]
\begin{center}
\begin{tabular}
{lc}
Parameter & Value \\
\hline
Star & G2V \\
Mass ($M_\odot$) & $1.16 \pm 0.06$ \\
Radius ($R_\odot$) & $1.05 \pm 0.03$ \\
Effective Temperature (K) & $5780 \pm 80$ \\
Age (Gyr) & $<$1.78 \\
Rotation Period (d) & $12.4 \pm 0.1$ \\
Limb darkening coefficients & $u_1 = 0.44 \pm 0.01^*$ \\
& $u_2 = 0.10 \pm 0.02^*$ \\
\hline
Planet & \\
Mass ($M_{Jup}$) & $2.45 \pm 0.014$ \\
Radius ($R_{star}$) & $0.138 \pm 0.001^*$ \\
Radius ($R_{Jup}$) & $1.41 \pm 0.02^*$ \\
Orbital period (d) & $1.4857108 \pm 2\times 10^{-7}$ \\
Semi--major axis ($R_{star}$)  & $5.738 \pm 0.005^*$\\
Semi--major axis (AU) & $0.028 \pm 0.007^*$\\
Inclination angle & $(89.0\pm 0.1)^{\circ *}$ \\
\end{tabular}
\caption{Stellar and planetary parameters, those marked by an asterisk are the result of this work.}
\label{param}
\end{center}
\end{table}

The star is orbited by a 2.45 Jupiter mass planet in a close orbit, with a period of 1.49 days, and semi--major axis of only $5.7 ~R_{star}$ \citep{Desert11}. The orbital inclination, $89^\circ$, is such that the planet eclipses the star very close to its equator, at a projected latitude of only $-4.6^\circ$ (here arbitrarily chosen to be in the Southern stellar hemisphere). During the 1240 days of observation by the Kepler mission, of 834 possible transits, only 583 had complete transit data. 
Most of the incomplete transits were due to data gap of the satellite, few of them had data only during part of the transit and were not considered in our analysis.

\begin{figure}[h]
\plotone{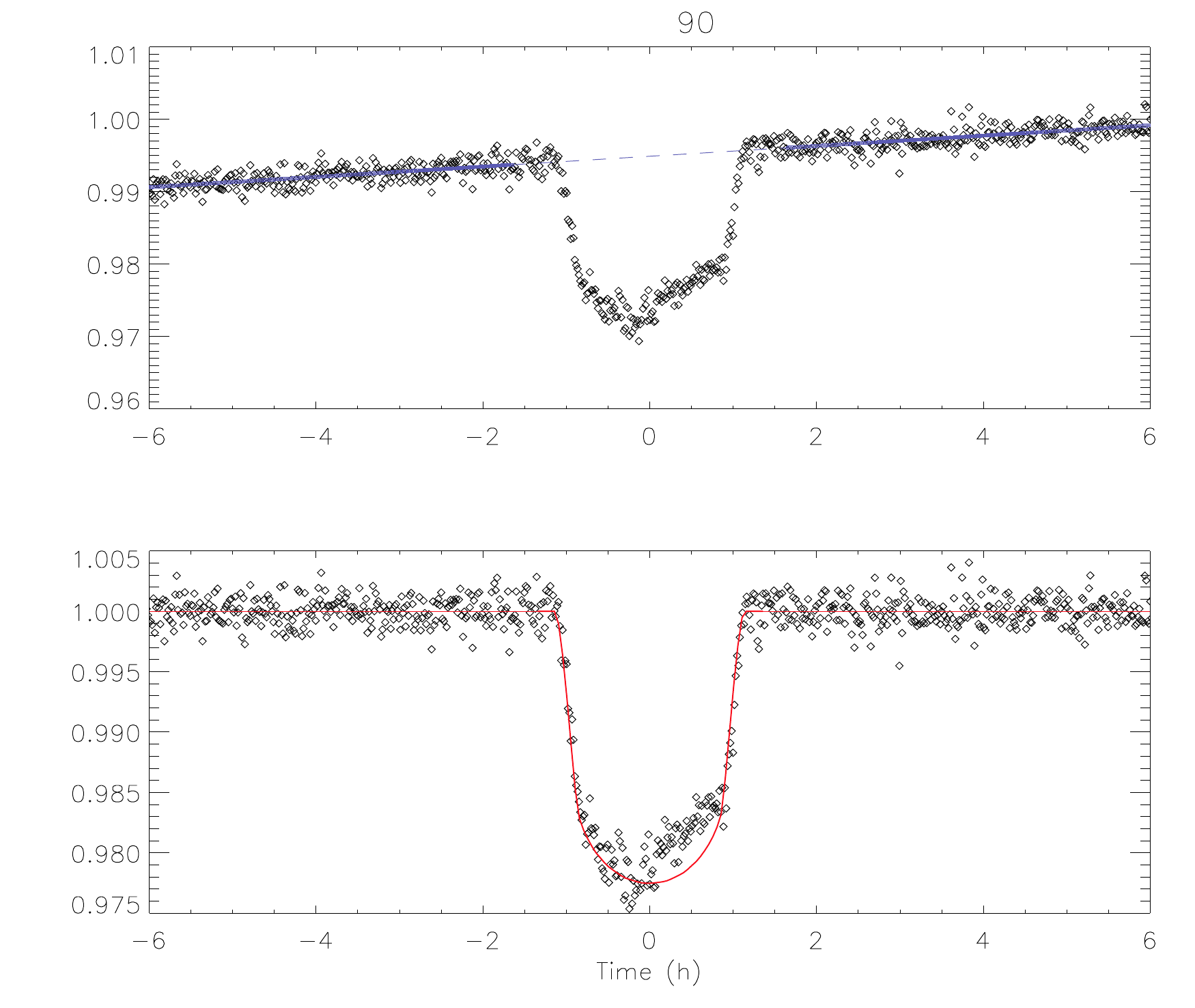}
\caption{{\it Top:} Light curve segment of the 90th transit of Kepler-17 b and the linear fit to the portions outside transit between 1.6 and 6 h (blue line). {\it Bottom:} The light curve data after the subtraction of the linear fit and normalisation to 1. The red curve represents the transit model of a spotless star.}
\label{cut_transit}
\end{figure}

Since we are interested in the light curve only during the transits, more specifically  plus or minus 6 hours from mid transit, it is not necessary to eliminate the overall rotational modulation or jumps between quarters. We simply do a linear fit to the data between 1.6 and 6 h, before and after mid transit, and then subtract it from the light curve. Next this transit light curve 12 h extract is normalised to 1. Figure~\ref{cut_transit} shows the data for the 90th transit, one of the cases with the largest slope, and the linear fit (blue dashed line), whereas the normalised and detrended light curve is shown in the bottom panel.

\begin{figure}[h]
\plotone{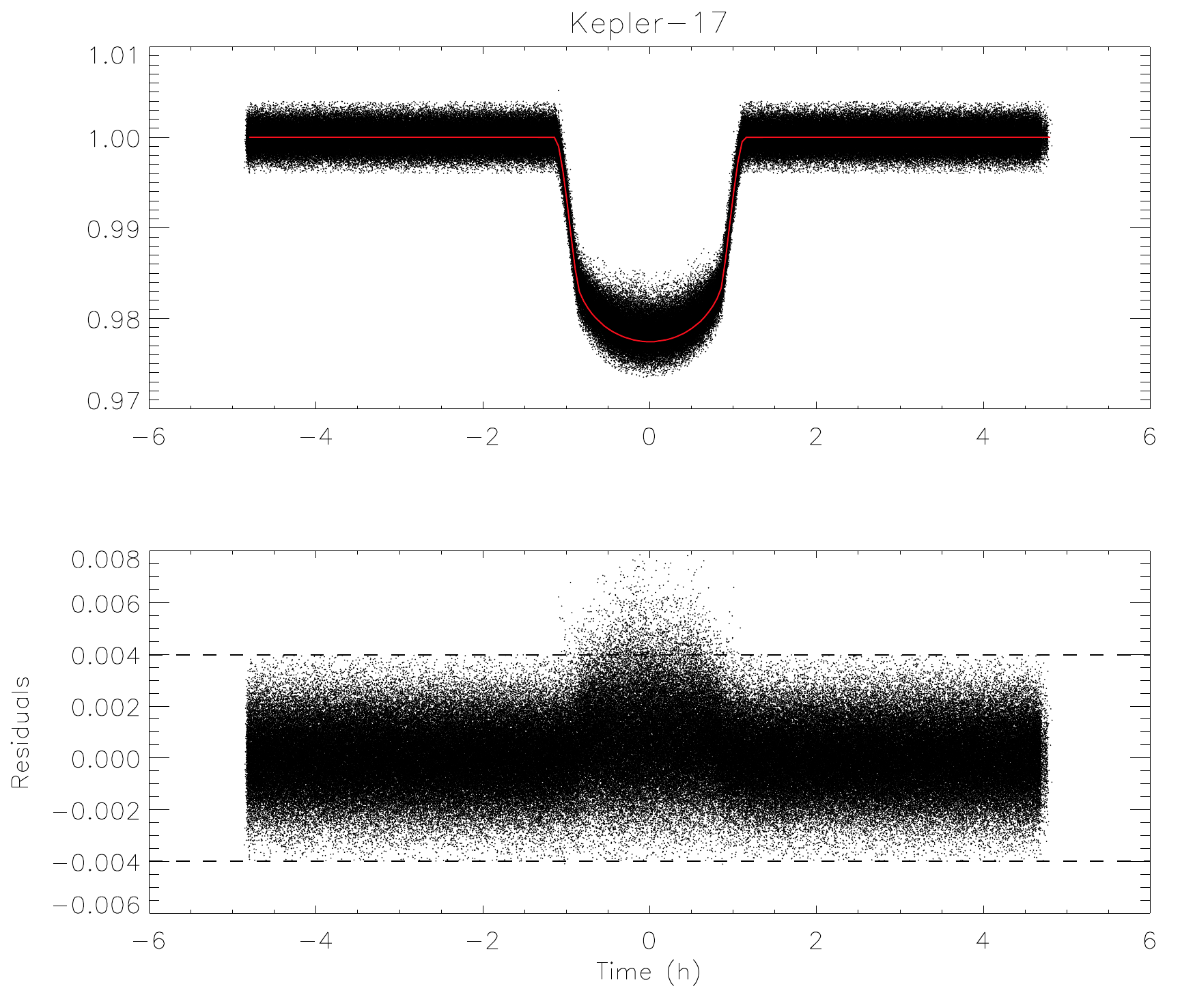}
\caption{{\it Top:} Folded light curve of Kepler-17 for a total of 583 transits. The red curve represents the transit model of a spotless star. {\it Bottom:} Residuals of transit light curves after subtraction of a spotless star model.}
\label{alltran}
\end{figure}

The folded light curve of all 583 valid transits is shown in the top panel of Figure~\ref{alltran}. Only the PDC\_SAP Short Cadence ($\sim 1$ min) Kepler light curve was used for a total of 16 quarters.

To obtain the spotless light curve, first the deepest transits were chosen, assumed to have very little or no activity (i.e. spots). Next the data from these 77 transits were binned into a single transit and fit by least chi-square minimisation routine (AMOEBA) using the model detailed in \cite{Silva03} but without any spots. The five parameter (planet radius, semi--major axis, inclination angle, and limb darkening coefficients) fit resulted in an increase in the radius of the planet and the semi--major axis of the orbit by about 7 and 8\%, respectively, with respect to the values reported by \cite{Desert11}. These parameters are given in Table~\ref{param}, and the resulting fit is shown as the red line on the bottom panel of Figure~\ref{cut_transit} and the top panel of Figure~\ref{alltran} applied to all the transits. 

The residuals of the subtraction of this model from all the data is plotted on the bottom panel of Figure~\ref{alltran}. The increase in the dispersion of the points during the transit is clearly seen and interpreted as being caused by the signatures of spots on the transit light curve. The horizontal line corresponds to 10 times the average CDPP (Combined Differential Photometric Precision, \cite{Christiansen12}) computed for all quarters. This value is used as a threshold for spot modelling, that is, we only fit the residuals that exceed this value.

\section{Starspot model}\label{Model}

\begin{figure}[h]
\plotone{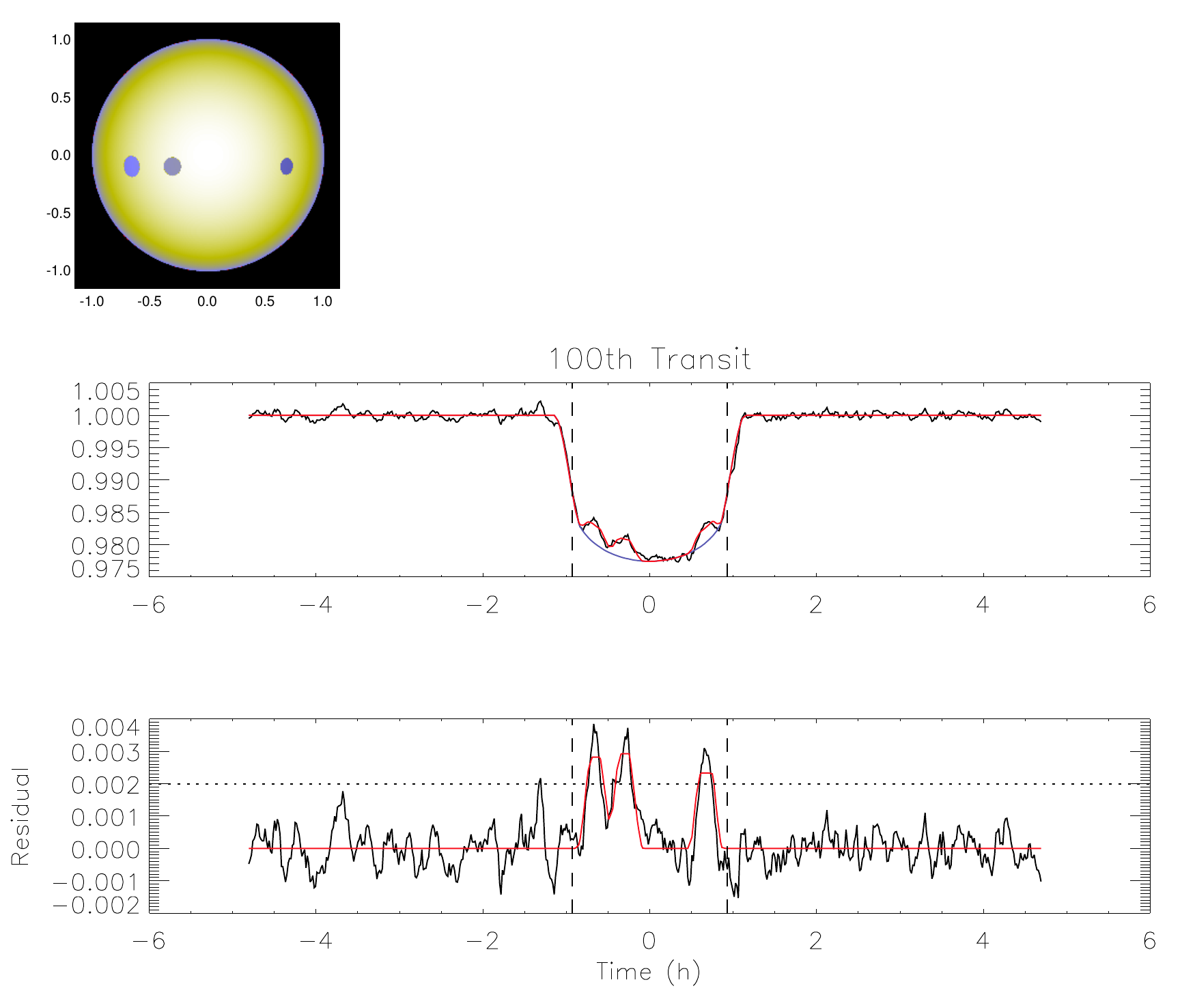}
\caption{The 100th transit of Kepler-17b was taken as a typical example of the spot fit by the model.
{\it Top:} Synthesised star with three spots. {\it Middle:} Transit light curve, over plotted are the model of a spotless star (blue) and a star with three spots  (red curve). {\it Bottom:} Residuals of transit light curve after subtraction of a spotless star model. The red curve shows the fit to the data ``bumps". The dotted horizontal line represents the threshold for spot modelling, and is 10 times the CDPP, whereas the vertical dashed lines represent the transit portion that is modelled within $\pm 70^\circ$ longitudes.}
\label{exfit}
\end{figure}

The model used here, described in \citet{Silva03}, simulates the crossing of a dark disc, the planet, in front of a synthesised star with limb darkening (see top panel of Figure~\ref{exfit}). The quadratic limb darkening law is given by:

\begin{equation}
 {I(\mu) \over I_c} = 1 - u_1 (1-\mu) - u_2 (1-\mu)^2
 \end{equation}
 
 \noindent where  $\mu = \cos(\theta)$, and $\theta$ is the angle between the line--of--sight and the emergent intensity. Thus, $I_c = I(\mu=1)$ is the intensity at the stellar disc center. The coefficients used for Kepler-17 are given in Table~\ref{param} and are very similar to those given by  \cite{Desert11} ($u_1=0.433$ and $u_2=0.101$) . 
  
 The orbit of the planet is then calculated according to the semi--major axis and inclination angle, and is assumed to be circular (i. e., zero eccentricity). Even though the model allows for the transit to be oblique, that is when the orbital plane is not parallel to the stellar equator, in the case of Kepler-17, following \citet{Desert11}, we considered the orbital axis to coincide with that of the stellar spin. Every two minutes (or the desired time), the dark disc of the planet is centred at its estimated position within the orbit. The sum of the intensity of all the pixels in the image yields the light curve value at that instant in time. 
 
 An unique  feature of this model is the possibility to add round spots to the stellar disc. Each spot is modelled with three parameters: radius (in units of planet radius, $R_p$), intensity (with respect to the central stellar disc intensity, $I_c$), and longitude, since the latitude is fixed and depends on the transit projection. The model also considers the foreshortening effect when the spots are close to the stellar limb. 
  
 An example of a transit across a star with spots is shown in Figure~\ref{exfit}. The top panel depicts the synthesised limb darkened star with three spots, whereas the middle panel shows the data (black curve), smoothed by a running mean every 5 points. Also shown in this panel is the model light curve of a star without any spots (blue curve).  Subtracting this model from the data yields the residuals, shown in the bottom panel of Figure~\ref{exfit}, where three ``bumps" that superseded the threshold are clearly seen. These are interpreted as signature of spots and are modelled as such.
 
To avoid overfitting, we only consider the signals that are above 10 times the average CDPP of the smoothed data, marked by the dotted horizontal line. Moreover, we only fit the spot signals within longitudes of $\pm 70^\circ$ to avoid the steep ingress and egress regions of the transit light curve. This region is delimited by the vertical dashed lines in both the middle and bottom panels of Figure~\ref{exfit}. 
From the 583 transits, only 507 transits satisfied these criteria.
The maximum number of spots in a given transit was set to 4, only for one transit was it necessary to fit 5 spots. 

The residuals of each transit light curve were fit individually using this model. The number of spots was determined {\it a priori} for each transit, as well as the initial guess for its longitude, $lg_{spot}$, calculated from the approximate time of the ``bump" maximum intensity, $t_s$ (in hours), given by:

\begin{equation}
lg_{spot} = {\rm asin} \left[{a \cos \left(90^\circ -  {360^\circ\ t_s \over 24\ P_{orb}}\right) \over \cos(lat_{spot})}\right]
\end{equation}

\noindent where $a$ is the semi--major axis and $P_{orb}$, the orbital period in days. Initial guesses for the radius and intensity of the spots were fixed at 0.5. The best fit was then calculated using the AMOEBA routine, that minimizes $\chi^2$. These fits are depicted as red curves in Figure~\ref{exfit}.

\section{Spot characteristics}\label{Spots}

From the 507 transits that showed ``bumps" in the residual light curve  above the adopted threshold, a total of 1069 spots were modelled, that amounts to an average number of $2 \pm 1$ spots per transit. Each spot was modelled by three parameters: size, intensity and longitude. The spot parameter errors estimated in a conservative manner are: approximately 10\% for radii and intensities and 5 degrees for the longitudes.

\begin{figure}[h]
\plotone{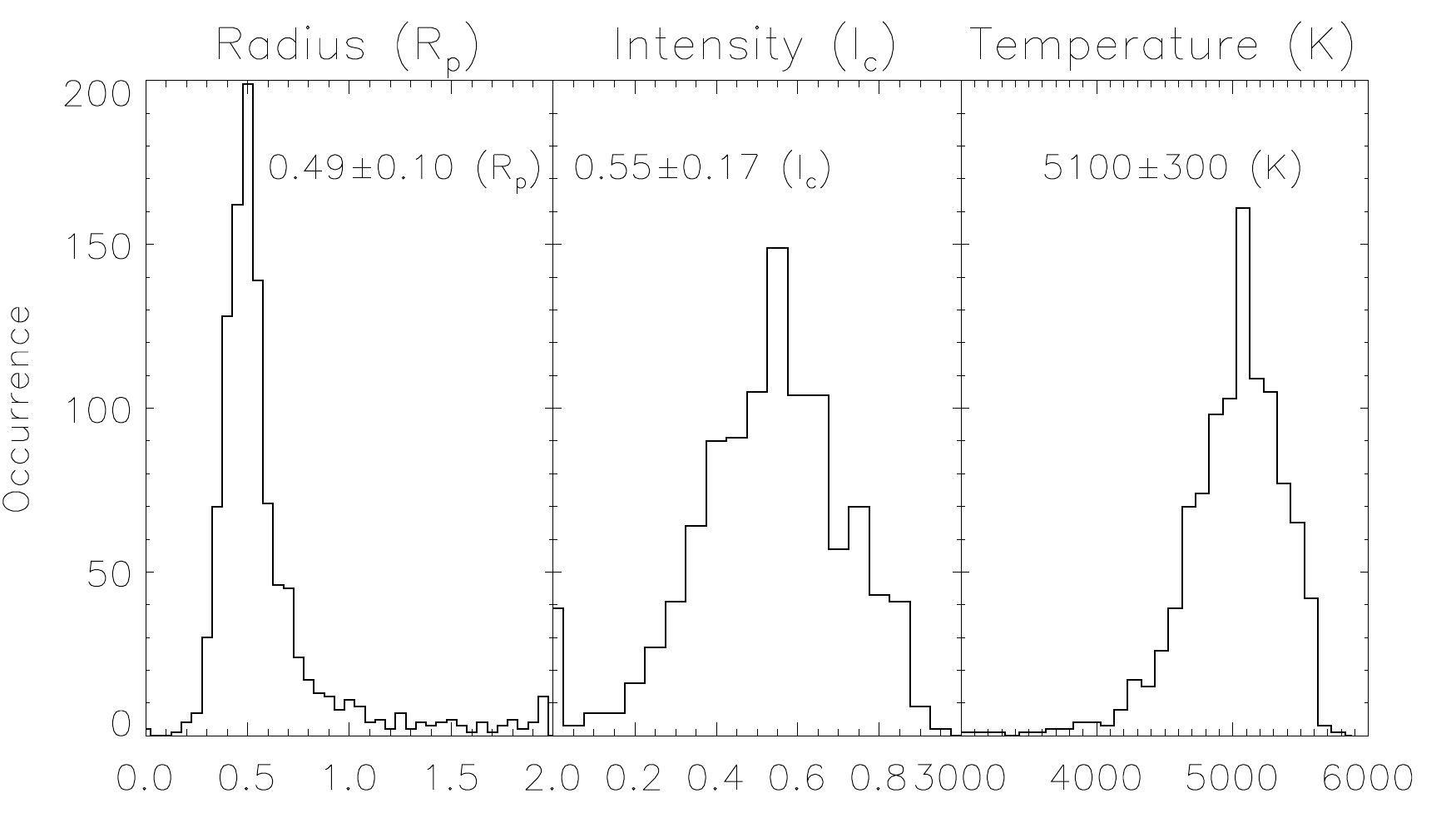}
\caption{Parameters distribution of the modelled 1069 spots: radius in units of $R_p$ (left), intensity in units of stellar central intensity, $I_c$ (middle), and temperature (right).}
\label{histo}
\end{figure}

Distributions of the spots parameters are shown in Figure~\ref{histo}, for the radius and intensity. The average radius of the spots on Kepler-17 is $0.49 \pm 0.10 R_p$, that is $49,000 \pm 10,000$ km. The spots intensity can be converted to temperature, if one assumes that both the stellar photosphere and the spot radiate as a blackbody (see \citet{SilvaValio10}). The rightmost panel of Figure~\ref{histo} shows the spots temperature, with an average value of $5100 \pm 300$ K. 

\begin{figure}[h]
\plotone{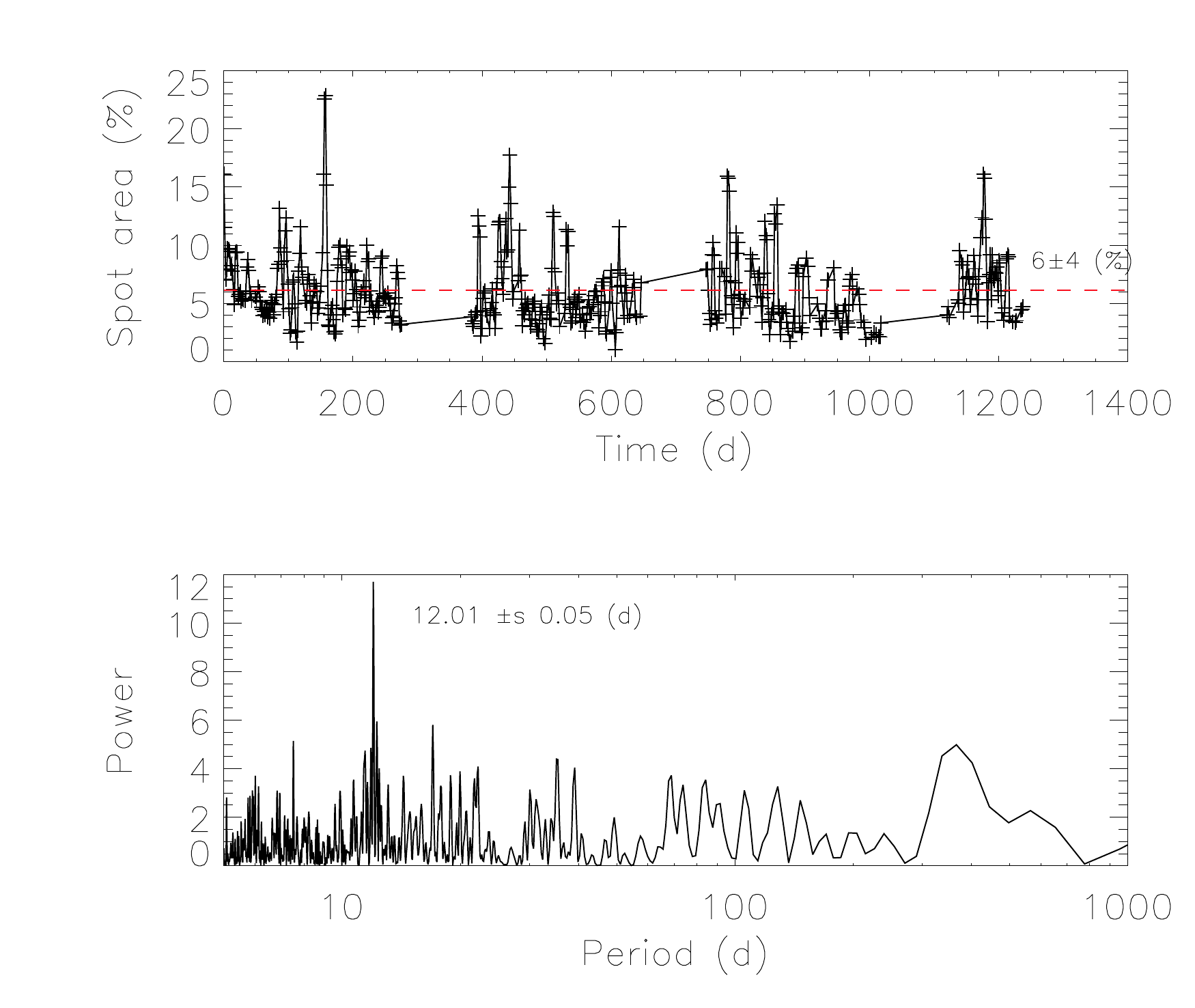}
\caption{{\it Top:} Stellar surface area covered by the spots for all the transits. The red dashed line represents the average value of $6 \pm 4$ \%. {\it Bottom:} Lomb-Scargle periodogram of the area, showing the peak at 12.0 days.}
\label{area}
\end{figure}

As  in \citet{SilvaValio10}, one can estimate the stellar surface area covered by the spots within the transit band. This was done for each transit and the percentage of spot area in time is shown in the top panel of Figure~\ref{area}, where the dashed horizontal line represents the average value of $6 \pm 4$ \%. This value is larger than the $<1$\% of the solar case. It is, however, much smaller than the value of 10--15\% found by \cite{Bonomo12a}. Since our model only considers spots close to the equator (latitude band of $0^\circ$ to  $-10^\circ$ ), this discrepancy may be an indication that the majority of  spots are located at higher latitudes. This is the case for sunspots.

A clear periodicity can be seen in the area coverage. A Lomb-Scargle periodogram performed on the area data yield a maximum periodicity of $12.01 \pm 0.05$ d, that is smaller than the average rotation period of 12.4 d, and equals very closely 8 times the orbital period of the planet. For a detailed analysis of the periodogram and the presence of a magnetic cycle of $1.12 \pm 0.16$ yr, see \cite{Estrela16}.

\begin{figure}[h]
\plotone{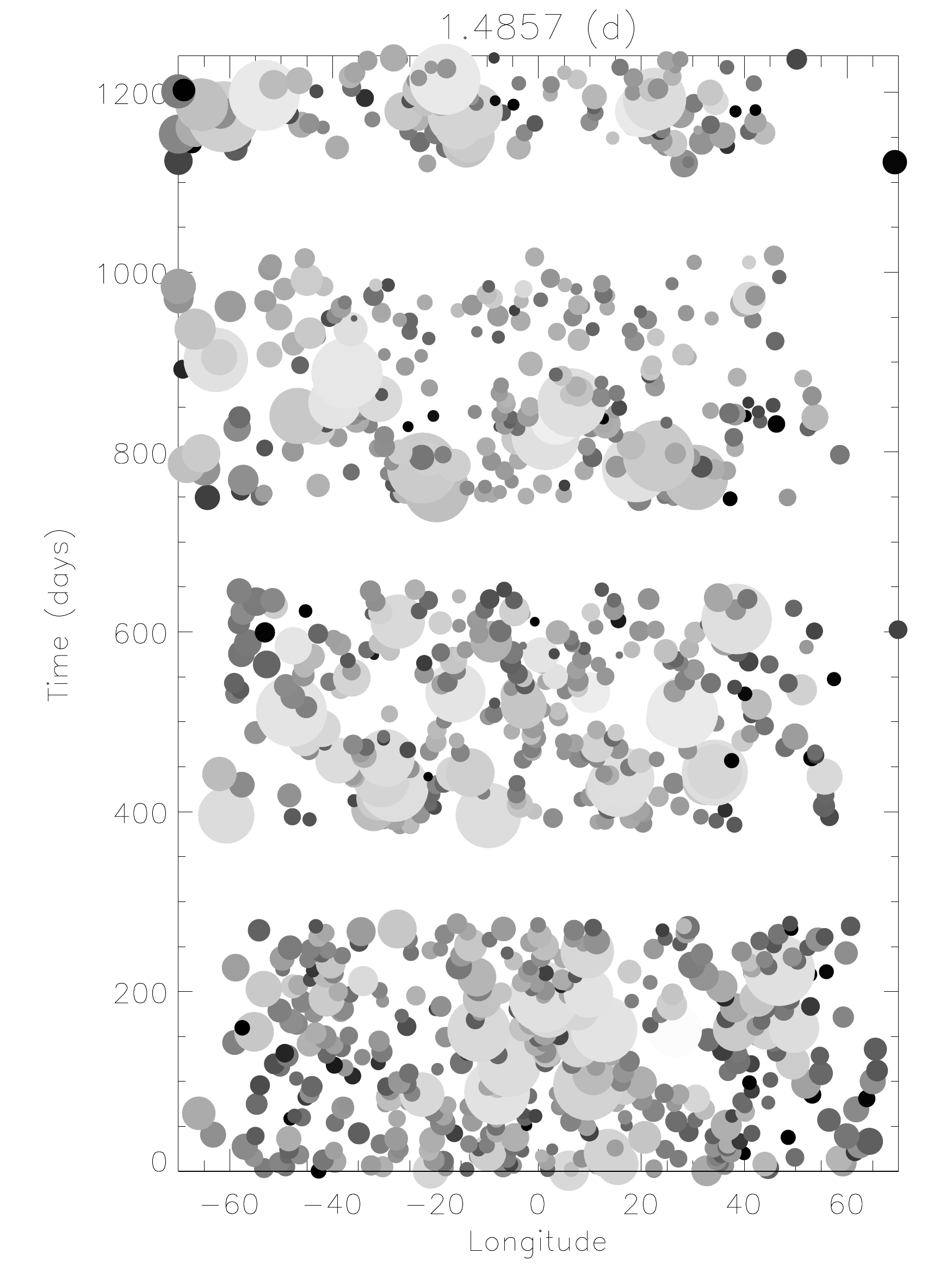}
\caption{Longitude map of the spots in time for all the transits. The relative size and color of the spots represents its radius and intensity. This map was constructed considering the view from Earth, so that the spots maximum (minimum) longitude is $70^\circ$ ($-70^\circ$).}
\label{mapporb}
\end{figure}

Figure~\ref{mapporb} shows the spots position (longitude), size, and intensity for each transit during the whole observing period. All the transits spotted bands are stacked vertically, with time increasing in the y--axis, whereas the spot longitude is given in the x--coordinate. This spot longitude is given with respect to a coordinate system fixed at Earth, $lg_{topo}$, where zero longitude corresponds to the central meridian of the star as seen from Earth, at mid transit time. Each spot is represented by a circle with relative size and intensity, darker spots having higher contrast. Note that only spots within $\pm 70^\circ$ longitudes were fit. We call this type of figure, a map of the stellar surface for each transit. 

\section{Rotation and differential rotation}\label{Res}

\begin{figure}[h]
\plotone{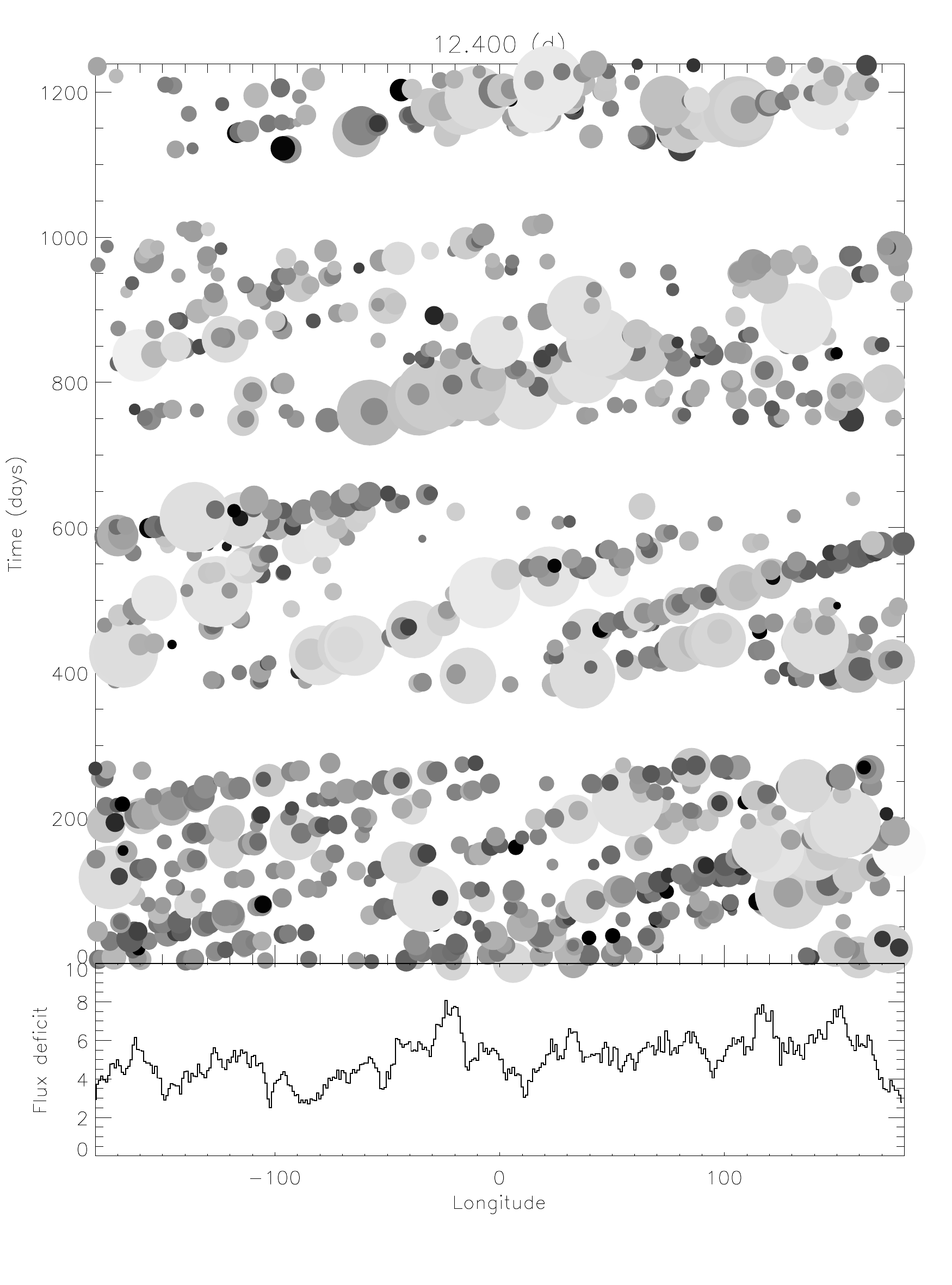}
\caption{{\it Top:} Longitude map of the spots in time for all the transits. This map was constructed considering a coordinate system that rotates with the average stellar period of 12.4 d. {\it Bottom:} Flux deficit, or the sum of the spots contrast for each longitude bin.}
\label{mapprot}
\end{figure}

Since we are interested in determining the rotation profile of Kepler-17, it is interesting to map the starspots onto the rotation frame of the star. This can be done by converting the longitude in a topocentric coordinate system, $lg_{topo}$, to one that rotates with the star, $lg_{rot}$:

\begin{equation}
lg_{rot} = lg_{topo} - 360^\circ {n P_{orb} \over P_{star}}
\label{lg}
\end{equation}

\noindent where $n$ is the transit number, $P_{orb}$ is the planetary orbital period and $P_{star}$ is the stellar rotational period.
A map of the spots on the surface of Kepler-17 in a rotating frame with a period of 12.4 d is shown in Figure~\ref{mapprot}. Note that here the longitudes vary from $-180^\circ$ to $+180^\circ$, zero longitude defined as that of the central meridian at the time of the first mid transit. The bottom panel of the figure depicts the flux deficit due to the spots (in arbitrary units), that is the total intensity subtracted from the stellar surface within a certain longitude due to the presence of spots.

\subsection{Rotation period at the transit latitude}

As can be seen from Figure~\ref{mapprot}, the spots seem to follow an inclined line. According to
 \citet{Valio13}, this indicates that the value of $P_{star}$ used in Eq.~\ref{lg} does not represent the correct one for the rotation period at that stellar latitude. The next step is to calculate the period that will straighten the trends of spots vertically (same longitude), thus attesting that we are detecting the same spot on a later transit. The procedure is the same as that described in  \citet{Valio13}, and applied in \citet{SilvaValio11}.
 
 \begin{figure}[h]
\plotone{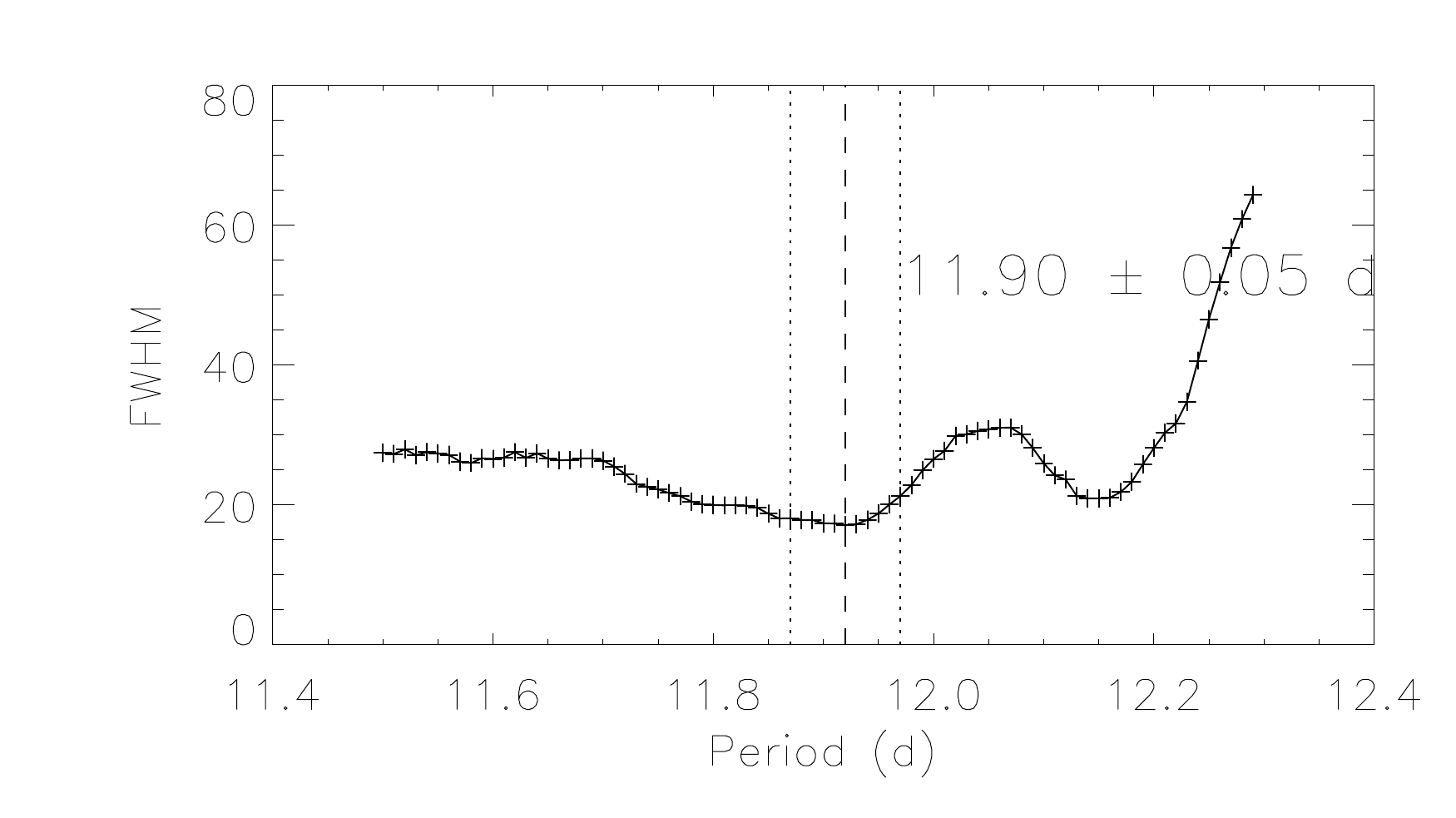}
\caption{FWHM of the auto-correlation function of the spot flux deficit calculated for maps with different stellar rotation periods. The rotation period at the latitude of the transit, $-4.6^\circ$ is taken as the smallest FWHM, that in this case equals 11.92 d.}
\label{fwhm}
\end{figure}

\begin{figure}[h]
\plotone{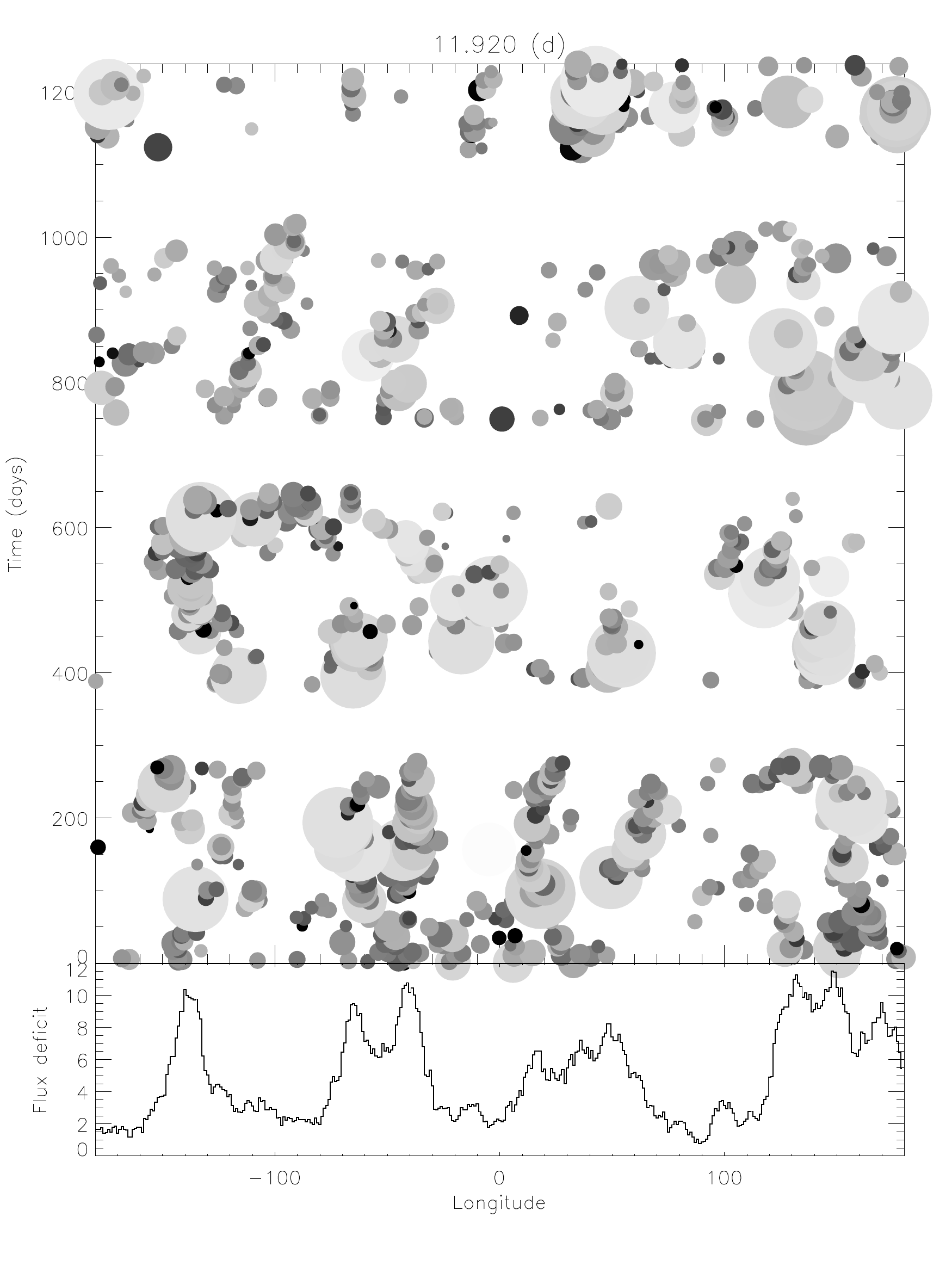}
\caption{Same as Figure~\ref{mapprot} but for a period of 11.92 d.}
\label{mapplat}
\end{figure}

 First, we vary $P_{star}$ and calculate the auto--correlation function for the flux deficit (bottom panel of Figure~\ref{mapprot}). Then, we calculate the full-width at half-maximum of the auto--correlation function of the flux deficit for different rotation periods, $P_{star}$. Figure~\ref{fwhm} shows the FWHM of this auto--correlation function with a running mean of 10 points. From these, we chose the period that corresponds to the thinnest auto--correlation function, or the minimum of the FWHM function presented on  Figure~\ref{fwhm}, which implies that the spots are vertically aligned.
 For this star, we found that the period that best describes the rotation of the latitude band centred at $-5^\circ$ is $11.92 \pm 0.05$ d (indicated by the vertical lines in Figure~\ref{fwhm}), somewhat shorter than the average out--of--transit period of 12.4 d. The spot longitude map that corresponds to the 11.92 d period is depicted in Figure~\ref{mapplat}, with the corresponding flux deficit function plotted in the bottom panel.

\subsection{Differential rotation}

\begin{figure}[h]
\plotone{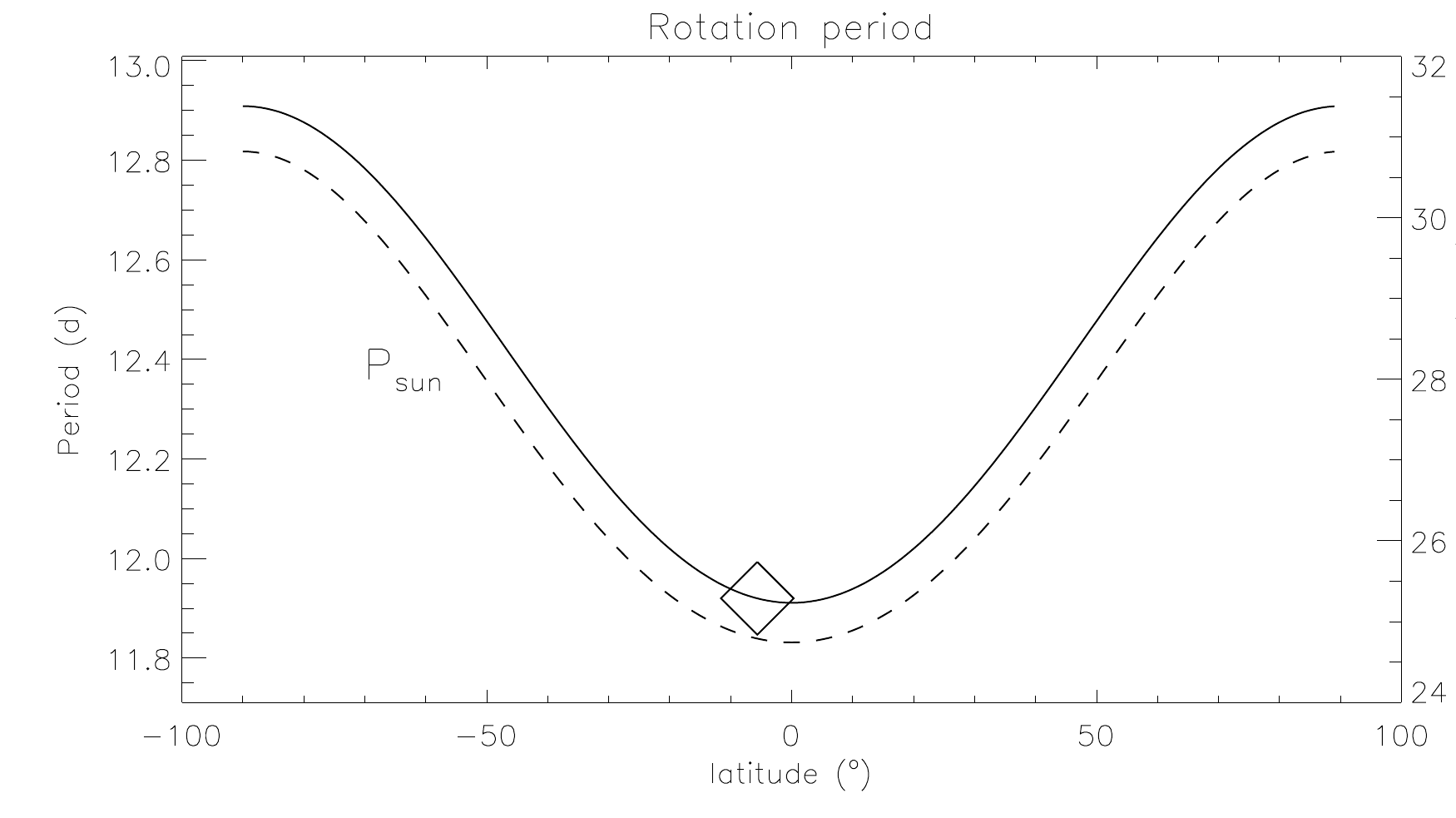}
\caption{Differential rotation profile, that is the rotation period as function of latitude for Kepler-17 (solid line) and the Sun (dashed line, right axis). Both profiles are assumed to vary as $sin^2(lat)$. The period determined in Figure~\ref{fwhm} is represented by the diamond.}
\label{difrot}
\end{figure}

Now we can estimate the differential rotation of the star, since we know that the average rotation period is 12.4 d, while the period close to the equator of Kepler-17 is shorter, 11.92 d. Since, we only measured the rotation period in one latitude, we need to assume a rotation profile for this star. Here we consider a solar--like profile, since the star is similar to the Sun:

\begin{equation}
 \Omega (\alpha) = \Omega_{eq} - \Delta\Omega ~\sin^2(\alpha)
 \label{eq:sunrot}
\end{equation}

\noindent where $\Omega$ is the angular velocity, $\Delta\Omega$ the rotation shear, and $\alpha$ is the stellar latitude. To estimate the angular velocity and the rotation shear, we use both the average period of $12.4 \pm 0.1$ d, and the period of $11.92 \pm 0.05$ calculated at $lat=-4.6^\circ$, following \cite{SilvaValio11}. The result of the fit is shown in Figure~\ref{difrot}, where the 11.92 d period at $lat=-4.6^\circ$ is identified by the diamond (with its size depicting the uncertainty in the period).

Using these values and their uncertainties, we estimated a rotation shear of $\Delta\Omega=0.041 \pm 0.005$ rd/d and an equatorial period of $P_{eq} = 2\pi / \Omega_{eq} = 11.91$ d. The shear is similar to the solar case of 0.050 rd/d, whereas the equatorial period is equal to 8.0 times the orbital period of the planet, implying that there is strong star--planet interaction going on in the Kepler-17 system. From these values, one can also calculate the relative differential rotation,  $\Delta\Omega/\bar\Omega=8.0 \pm 0.9$ \%, where $\bar\Omega=2\pi/P_{star}$ is the average rotation rate.
 
\section{Conclusions}\label{Concl}

We have modelled the light curve of star Kepler-17, using the spot model described in \citet{Silva03}, for 583 transits where a total of 1069 spots were identified well above the noise level. Assuming the detection of the same spot on a later transit, we were able to infer the rotation period of the star at the latitude band ($\sim 5^\circ$) occulted by the planetary transit. The spot  transit data were best fit by a period of $11.92 \pm 0.05$ d, as opposed to the out--of--transit average period of $12.4 \pm 0.1$ d.

Considering a rotation profile with latitude similar to that of the Sun, given by Eq.~\ref{eq:sunrot}, it was possible to determine the differential rotation, or shear, $\Delta\Omega = 0.041 \pm 0.005$ rd/d. This value is close to the solar value of 0.050 rd/d. Kepler-17 is a solar--like star of spectral type G2V, with the same effective temperature as the Sun, 5781 K \citep{Bonomo12b}. According to \citet{Barnes05}, the differential rotation of a star increases with its effective temperature (see their Figure 2). Thus it is not surprising that Kepler-17 shear is approximately the same as the solar one.

Nevertheless, Kepler-17 is a younger star with less than 1.78 Gyr, and thus spins faster, every 12 days or so, against the 27 d mean rotation period of the Sun. It is also more active than our star, with an average surface area coverage of 6\%, whereas on the Sun it is less than 1\%. 

Magnetic dynamos of stars are thought to be generated on the bottom of the convective zone by the $\Omega$-effect, and amplified as the flux tubes rises by the $\alpha$-effect pumped by the Coriolis force from the stellar rotation. Therefore, it appears that faster rotation plays a more important role in the generation of the magnetic field of Kepler-17, than the shear.

\acknowledgments

We are grateful to the Brazilian funding agency FAPESP ($\#$2013/10559-5) for partial financial support of this work.

\clearpage


\end{document}